\documentstyle[prl,aps,epsfig] {revtex}
\begin{document}
\advance\textheight by 0.2in
\draft
\twocolumn[\hsize\textwidth\columnwidth\hsize\csname@twocolumnfalse%
\endcsname
\title{On-off Intermittency in Stochastically Driven Electrohydrodynamic 
Convection in Nematics}
\author{Thomas John$^{1,2}$, Ralf Stannarius$^2$, and Ulrich Behn$^1$}
\address{$^1$Institut f\"ur Theoretische Physik and $^2$Institut f\"ur
	 Experimentelle Physik I, Universit\"at Leipzig, \\ P.O.B.  920, D-04009
	 Leipzig, Germany}
\date{February 12, 1999}
\maketitle
\begin{abstract}

We report on-off intermittency in electroconvection of nematic liquid crystals
driven by a dichotomous stochastic electric voltage.  With increasing voltage
amplitude we observe laminar phases of undistorted director state interrupted by
shorter bursts of spatially regular stripes.  Near a critical value of the
amplitude the distribution of the duration of laminar phases is governed over
several decades by a power law with exponent $-3/2$.  The experimental findings
agree with simulations of the linearized electrohydrodynamic equations near the
sample stability threshold.

\end{abstract}
%
] 

Systems at a threshold of stability driven by a stochastic or chaotic process
coupling multiplicatively to the system variables may exhibit on-off
intermittency characterized by specific statistical properties of the
intermittent signal.  Quiescent (or laminar) periods (off-states) are
interrupted by bursts of large variation (on-states); the duration of laminar
periods is governed by power laws with exponents universal over a broad class of
different systems.  Early theoretical studies considered systems with few
degrees of freedom modeled by differential equations \cite{Fujisaka} and
mappings \cite{Platt}.  There is increasing interest in systems with many
degrees of freedom \cite{Fujisaka97}, described by random map lattices
\cite{Yang}, larger systems of coupled nonlinear elements \cite{Fujisaka98}, and
partial differential equations \cite{Fujisaka98,Fukushima98}.  Experimental
results are available mainly for nonlinear electric circuits \cite{electr};
on-off intermittency was also observed in a spin wave experiment
\cite{Roedelsperger95}, in optical feedback \cite{Sauer96}, and a gas discharge
plasma system \cite{Feng98}.  Here we first report about on-off intermittency in
a spatially extended dissipative system, viz.  electroconvection (EC) in nematic
liquid crystals driven by a stochastic voltage.

EC in planarly aligned nematics is a standard system for pattern formation, for
recent reviews see, e.g.  \cite{Reviews}.  In the presence of an electric field
$E$ a spontaneous fluctuation of the director leads due to the anisotropic
conductivity to a formation of space charges which tend to destabilize the
homogeneously ordered state.  With increasing strength of the driving field one
observes a hierarchy of convection patterns of increasing complexity.  The
patterns depend on external parameters such as amplitude, frequency and wave
form of the driving voltage which are conveniently adjustable in the experiment.
The hydrodynamic flow induces a modulation of the director field and thus of the
effective indices of refraction which leads to transmission patterns easily
observed with a microscope.

In previous experiments, the superposition of a deterministic AC field with a
stochastic field, $E=E^{det}(t)+E^{stoch}(t)$, was studied.  A variety of
noise induced phenomena including stabilization or destabilization of the
homogeneous state, and a change from continuous to discontinuous behaviour of
the threshold as a function of the noise strength was observed
\cite{Kai79,Brand85,Kai87,Kai89b,Amm97} and has stimulated theoretical work
\cite{Behn85,Behn98}.  

As long as the characteristic time of the noise $\tau_{\text{stoch}}$ is small
compared with characteristic times of the system the threshold towards pattern
formation appears sharp as for deterministic driving.  It is however typical for
EC in nematics that one of the systems characteristic times decreases both with
increasing strength of the threshold voltage and with increasing wave number of
the pattern and may reach the order of $\tau_{\text{stoch}}$ \cite{Behn98}.  In
this case, one observes intermittent bursts of a regular spatial stripe pattern
which makes a na\"{\i}ve experimental determination of the stochastic threshold
difficult \cite{Amm97}.  A similar phenomenon was noticed in a highly doped
nematic material for very high strength of the noise \cite{Brand85} where a
direct transition towards chaos occurs via intermittent bursts of spatially
incoherent structures embedded in a homogeneous background.

In this paper, we consider the simplest case of pure stochastic excitation,
$E=E^{stoch}(t)$.  To achieve a statistical characterization we have determined
experimentally the distribution of the duration $\tau$ of laminar, i.e.
undistorted phases (off-states) which are interrupted by bursts of a stripe
pattern (on-states).  Approaching a critical voltage from below, this
distribution is governed by a power law $\tau^{-3/2}$ over several orders of
$\tau$ as it is typical for on-off-intermittency.  This result is confirmed by
simulations of the linearized electrohydrodynamic equations at the sample
stability threshold \cite{Behn98}.

We use the standard experimental set-up with a commercial cell (Linkam)
providing planar anchoring of the director by antiparallely rubbed polyimide
coatings, two transparent ITO electrodes of 5$\times$5mm$^2$ and a cell gap of
$d=50\mu$m.  The nematic material is a mixture of four disubstituted
phenylbenzoates \cite{Amm97} which has a nematic range from below room
temperature to 70.5$^\circ$C.  The cell temperature is controlled at
$30^\circ$C by a Linkam heating stage.  Images are recorded using a Jenapol-d
polarizing microscope and a Hamamatsu B/W camera with controller C2400.  The
transmission images of light polarized along the director easy axis $\vec n_0$
are recorded digitally.  We resolve the images in 500$\times $400 (330$\times
$250) pixels with 8 bits greyscale and calculate standard deviations in real
time at a rate of 1/7 (1/20) s in the conductive (dielectric) regime.

As driving process we use the dichotomous Markov process (DMP) which is easily
generated and faciliates the theoretical analysis.  The DMP $E_t^{\text{DMP}}$
jumps randomly between $\pm E$ with average rate $\alpha$.  The distribution of
times $\tilde{\tau}$ between two consecutive jumps is $\alpha
\exp{(-\alpha \tilde{\tau})}$;  the autocorrelation decays exponentially,
$<E_t^{\text{DMP}}E_{t'}^{\text{DMP}}> = E^2 \exp[-2 \alpha (t-t')]{} $, i.e.
$\tau_{\text{stoch}}=1/2 \alpha$.  We will refer to $\nu=\alpha /2$ as the mean
frequency and to $E$ as the amplitude of the noise.  Both in experiment and
simulation, sequences of the DMP are generated by the same algorithm;
technically $\tilde{\tau}$ is limited to vary between $10^{-4}$s and $10^4$s.  

\begin{figure} 
\psfig{figure=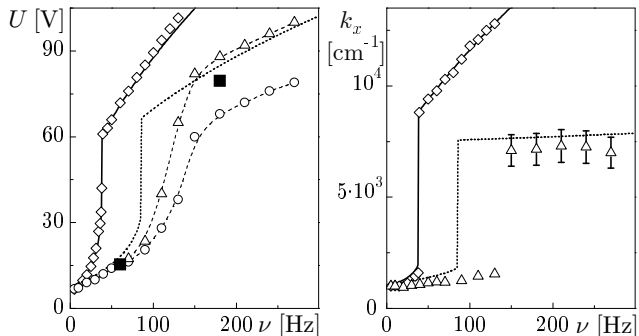} 
\caption{
\label{fig1}
Threshold voltages and wave numbers for driving with periodic and stochastic
square waves of (mean) frequency $\nu $.  Experimental data (diamonds:  periodic
case; circles, triangles, and full squares:  stochastic case, see text; dashed
lines guide the eyes) are compared with results from the two-dimensional theory
(full line:  periodic case; dotted line:  stochastic case).  } 
\end{figure}

For excitation by a periodic square wave, one finds a typical frequency
dependent threshold voltage $U_c$ ($U=Ed$) for the stability against formation
of normal rolls, cf.  Fig.  \ref{fig1}.  There is a sharp transition at
$\nu_c=38$ Hz between the conductive regime (oscillating space charges)
characterized by a wavenumber $k_x\approx\pi/d$ and the dielectric regime
(oscillating director deflections), where $k_x$ is an order of magnitude larger,
cf.  Fig.  \ref{fig1}.
For stochastic driving, we have observed bursts of stripe pattern uniform across
the system already below the onset threshold for a periodic voltage of the same
frequency.  The stochastic voltage has a broad frequency spectrum which contains
low frequency contributions.  Occasional bursts of convective pattern can be
expected at voltages above the DC threshold ($\nu=0$ in Fig.  \ref{fig1}a).
With increasing voltage the frequency of the bursts increases.  In Fig.
\ref{fig1}a we have plotted the voltages which correspond to a ratio of 75\%
(circles) and 25\% (triangles) of laminar phases, respectively.  The full
squares at $\nu=60$ Hz and 180 Hz indicate the voltages for which the
experimentally determined distribution of laminar periods $\tau$ is best
described by a $\tau^{-3/2}$ law (see below).  The theoretical results obtained
from the sample stability criterion explained below agree very well with the
experimental data.  (Both in the periodic and the stochastic case we have used
the same material parameters.)
\begin{figure}[t] 
\psfig{figure=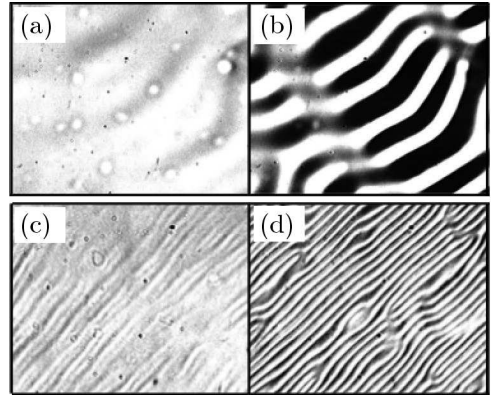} 
\vspace{1mm}
\caption{ \label{fig2}
Snapshots of the bursts of roll patterns (a, b) in the conductive regime
($\nu=60$ Hz, $k_x=1.2 \times 10^3$ cm$^{-1}$, area size 200$\times 164
\,\mu$m$^2$) and (c, d) in the dielectric regime ($\nu=180$ Hz, $k_x=7.2\times
10^3$cm$^{-1}$, area size 134$\times 102 \,\mu$m$^2$) at different times.  (a,
c) show the patterns just at the threshold of the on-state
($\sigma_{\text{rel}}=0.1$) whereas in (b, d) the rolls are fully developed
($\sigma _{\text{rel}}=0.8$).  }
\end{figure}
\vspace{-1mm}
The set of images in Fig.  \ref{fig2} shows the stripe pattern at different
times of a burst just at the threshold of the on-state and the fully developed
pattern in both the dielectric and the conductive regime.  We characterize the
intensity modulation of these patterns by $\sigma_{\text{rel}} =(\sigma
-\sigma_0)/(1 -\sigma_0)$ where $\sigma$ is the normalized standard deviation
from the average intensities taken over all pixels of the image at a given
instant, and $\sigma_0$ is the value of $\sigma $ for zero voltage.  This
procedure allows a real time characterization of the patterns.  It is justified
since the largest Fourier coefficient dominates the intensity modulation , and
both quantities have nearly equivalent traces.  Although the relation between
$\sigma_{\text{rel}}$ and the director deflections is nonlinear, the approach
used here is sufficient if we are interested only in the frequency and duration
of the bursts and not primarily in their amplitudes.

\begin{figure}[t] 
\psfig{figure=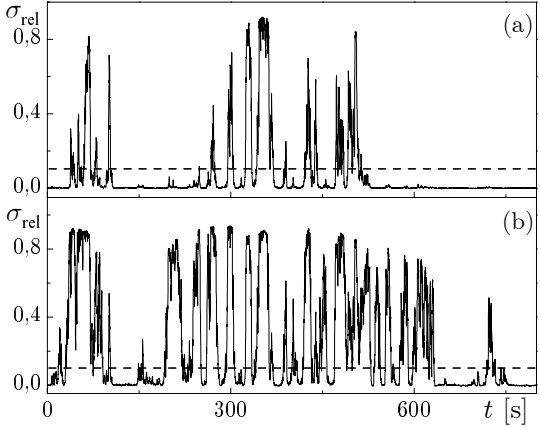} \vspace{1mm}
\caption{\label{fig3} 
Bursts of the intensity modulation $\sigma_{\text{rel}}(t)$ in the conductive
regime (a) just at the threshold and (b) 0.3 V above threshold for identical
trajectories of the DMP which makes about $9\times 10^4$ jumps in the 
period shown ($\nu=60$ Hz). The dashed lines indicate $\sigma_{lam}$ above which
the system is in the on-state and else in the off-state.} 
\end{figure}

In Fig.  \ref{fig3} we present trajectories of $\sigma_{\text{rel}}(t)$
observed in the experiment for different noise amplitudes but the same seed of
the driving DMP.  As laminar phase we define the period in which
$\sigma_{\text{rel}}(t)$ is below a threshold $\sigma_{\text{lam}}=0.1$ where
the choice of $\sigma_{\text{lam}}$ is not crucial.  The laminar periods are
interrupted by bursts of the convection structures, the frequency of which
increases with the applied voltage amplitude $U$.

\begin{figure}[p]
\psfig{figure=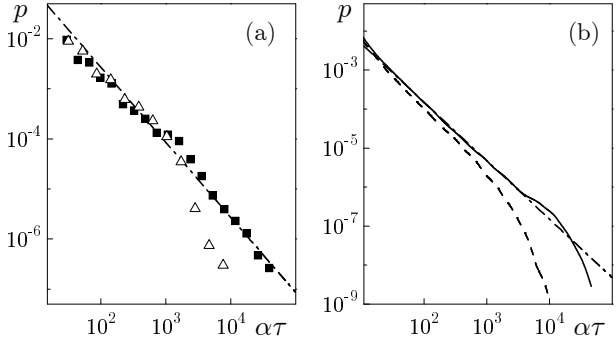}\vspace{1mm}
\caption{
\label{fig4}
Normalized distribution $p(\tau)$ in the conductive regime ($\nu$ = 60 Hz) just
at the stability threshold and slightly above.  Shown are results (a) from 
experiment  for $U=14.6$ V (squares) and $15.3$ V (triangles), and (b) from
simulations  for $U=U_c=18.2$ V (full line) and $U=19.0$ V (dashed line). 
Mode selection gives a wave number $k_x = 1484$ cm$^{-1}$ used in the 
simulation. The dash-dotted lines indicate a power law $\tau^{-3/2}$.  } 
\end{figure}

\begin{figure}[t]
\psfig{figure=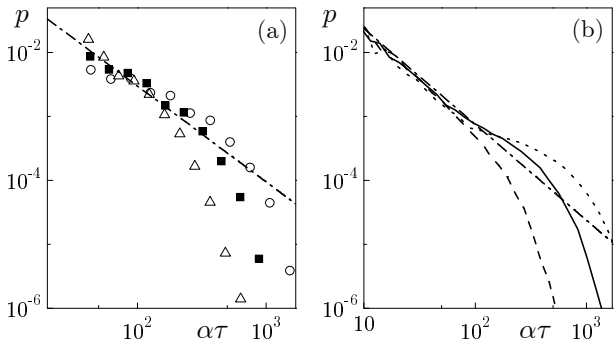} \vspace{1mm}
\caption{ \label{fig5}
Normalized distribution $p(\tau)$ in the dielectric regime ($\nu$ = 180 Hz)
slightly below $U_c$, just at $U_c$, and slightly above.  Shown are (a) experimental
results  for $U=76.0$ V (circles), 79.5 V (squares), and 83.0 V (triangles),
and (b) simulations  for $U=82.0$ V (dotted line), $U=U_c=84.9$ V (full line),
and $U=90.0$ V (dashed line).  Mode selection gives $k_x = 7670$ cm$^{-1}$ used
in the simulation.  Dash-dotted line as in Fig.  \protect{\ref{fig4}}.  }
\end{figure}

In Figs.  \ref{fig4} and \ref{fig5} we compare the experimentally determined
distributions $p(\tau)$ for the occurence of laminar phases of duration $\tau$
with those obtained in simulations in the conductive and the dielectric regime,
respectively.

\vspace{-6mm}
\vspace{-5mm}
\noindent 
The experimental histograms contain data from about 2500 bursts
of $\sigma_{\text{rel}}(t)$ recorded over $1 ...  6$ h.  With increasing
voltage amplitude the frequency of bursts increases, i.e.  longer laminar
periods occur less frequently.  At a critical voltage $U_c$ (full squares in 
Fig. \ref{fig1}a) the distribution is
governed by a power 
law $p \sim \tau^{-3/2}$ over several decades.  Deviations
occur for very small $\tau$ due to the finite time resolution and for very
large $\tau$ due to the background fluctuations from thermal noise.  Increasing
the amplitude beyond the critical voltage leads to exponential corrections to
the power law discussed below.  Increasing the voltage further, the system does
not reach a state of permanent convection, instead the patterns become more and
more spatially irregular while keeping its intermittent character.

The theoretical description of the laminar phases is based on the linearized
nemato-electrohydrodynamic equations.  In a two-dimensional idealization using
stress-free boundary conditions, inserting as a test mode a periodic roll
pattern characterized by wave numbers $k_x$ and $k_z = \pi/d$ they reduce to the
ordinary differential equations \cite{Behn98}

\begin{equation}\label{2}
\dot{\vec z} = {\bf C}(t)\vec z , \;\;\; \vec z = {\left( q \atop \psi \right)},
\end{equation}
where $q$ and $\psi$ are the amplitudes of the space charge density and 
of $\partial_x \varphi$ ($\varphi$ is the angle  between director and
electrode plates), and

\begin{equation}\label{3} {\bf C}(t) = -\pmatrix{ 1/T_q &\sigma_H
E_t^{\text{DMP}}\cr a E_t^{\text{DMP}}
&\Lambda_1-\Lambda_2 E^2\cr} .  \end{equation} 
The parameters $T_q$, $\sigma_H$, $a$, $\Lambda_1$, and $\Lambda_2$ depend on
material properties and on the wave number $k_x$ which is determined by
minimizing the threshold voltage, cf.  \cite{Behn98}.  Between two consecutive
jumps of $E_t^{\text{DMP}}$ at $t_{\nu}$ and $t_{\nu+1}$ the matrix ${\bf C}(t)$
is constant, and the time evolution is given by $\vec z(t)={\bf
T}^{s_{\nu}}(t-t_{\nu})\vec z(t_{\nu})$ for $t_{\nu} < t < t_{\nu+1}$ where
${\bf T}^{s}(t)$ is the time evolution matrix for
$\text{sgn}E_t^{\text{DMP}}=s$.  Iteration gives the formal solution
\cite{Behn98} for a given realization of the driving process with jumps at the
random times $t_{\nu},\, \nu =1,...,n $

\begin{equation}\label{4} 
\vec z(t)={\bf T}^{s_n}(t-t_n)\,\cdots\,{\bf T}^{s_0 }
(t_1-t_0)\vec z(t_0).  
\end{equation} 
The threshold voltage $U_c$ for a given wave number is determined by the zero of
the largest Lyapunov exponent $\lambda_1$ of the product of random matrices in
(\ref{4}) which can be calculated analytically as well as the second Lyapunov
exponent $\lambda_2 < \lambda_1$ at the threshold \cite{Behn98}.
Whereas $\tau _{\text{sys},1}= 1/|\lambda_1|$ diverges at the threshold we found
for the examples presented here in the conductive (dielectric) regime $\tau
_{\text{sys},2} = 1/|\lambda_2| = 3.0\times 10^{-3}$ s ($0.9 \times 10^{-3}$ s)
which is of the same order as $\tau_{\text{stoch}}=1/2\alpha=4.2\times 10^{-3}$
s ($1.4\times 10^{-3}$ s).

The numerical simulation generates trajectories $\vec z (t)$ starting from a
small nonzero initial value $\vec z (t_0)$, cf.  Eq.  (\ref{4}).  To model the
background of thermal fluctuations of $\psi$ we introduced a lower cutoff
$\psi_{\text{min}}$, i.e.  $\psi \rightarrow \psi_{\text{min}} \text{sgn} \psi $
for $|\psi | < \psi _{\text{min}}$.  In the dielectric regime we additionally
reset $q$ in a similar way.  A trajectory is considered laminar as long as
$|\psi |$ is smaller than a given threshold $\psi _{\text{lam}}=2\times 10^3\psi
_{\text{min}}$.  At $U_c$ the distribution is a power law $\tau^{-3/2}$ over
several orders of $\tau$ with deviations for very small and very large $\tau$ as
in the experiment, cf.  Figs.  \ref{fig4}b and \ref{fig5}b.  The range of
validity of the power law increases when $\psi _{\text{min}}/\psi _{\text{lam}}$
is lowered.  Also for voltages smaller or larger than $U_c$ the shape of the
simulated distributions is very similar to that obtained in experiment.  The
shoulder for large $\tau $ becomes more pronounced for voltages below the
critical voltage.

\begin{figure} 
\psfig{figure=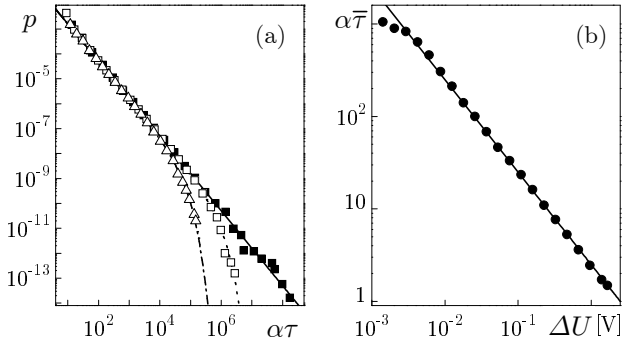} 
\caption{ \label{fig6}
(a) Normalized distribution $p(\tau)$ in the conductive regime ($\nu$ = 60 Hz)
obtained from simulations with very small lower cutoff at $U_c$ (full squares),
and slightly above:  $\Delta U=0.05$ V (squares) and $\Delta U=0.2$ V
(triangles).  The lines indicate the modified power law mentioned in the text.
(b) shows the mean value of duration of laminar periods $\bar{\tau}$ as function
of $\Delta U$.  $k_x$ as in Fig.  \protect{\ref{fig4}}.  }
\end{figure}

To distinguish between effects due to thermal fluctuations and due to a
deviation from the stability threshold $\Delta U=U-U_c$ we have also performed
simulations with $\psi_{\text{min}}=10^{-300}\psi_{\text{lam}}$ (which is
obviously too small for comparison with experiment).  At $U_c$ the power law
holds now over 8 decades.  Above threshold the results agree very well with $p
\sim \tau ^{-3/2}\exp(-$const$ \,\Delta U^2 \tau)$, see Fig.  \ref{fig6}a.  The
mean duration of laminar periods behaves like $\bar{\tau}\sim \Delta U^{-1}$,
cf.  Fig.  \ref{fig6}b.  Similar laws have been obtained analytically in
\cite{Platt} for a one-dimensional mapping.

We have found on-off intermittency in a spatially extended dissipative system
driven by multiplicative noise at parameter values where the first instability
is towards spatially regular structures.  If the strength of the noise
approaches a critical value both experiment and simulations of the
electrohydrodynamic equations lead to a power law with exponent $-3/2$ for the
distribution of laminar periods.  The simulations show that this critical value
is just the threshold of stability according to the sample stability criterion
\cite{Behn98}.

This study was supported by the DFG with Grants BE 1417/3 and STA 425/3.  We
acknowledge valuable discussions with Prof. Agnes Buka and Heidrun Sch\"uring.
\small 
\bibliographystyle{plain} 

\end{document}